\documentclass[aps,prl,twocolumn,groupedaddress]{revtex4}
\usepackage{graphicx}
\usepackage{amsmath,amsfonts,amssymb}
\usepackage{color}
\usepackage{float}
\begin{document}

\title{Mini-Superspace Universality \\
and No-Scale Quantum Cosmology}

\author{Tomer Ygael}
\author{Aharon Davidson}
\email{davidson@bgu.ac.il}
\homepage{http://www.bgu.ac.il/~davidson}
\affiliation{Physics Department, Ben-Gurion
University of the Negev, Beer-Sheva 84105, Israel}

\date{\today}

\begin{abstract}
We prove that, at the mini superspace level, and for
an arbitrary Brans-Dicke parameter, one cannot tell
traditional Einstein-Hilbert gravity from local scale
invariant Weyl-Dirac gravity. 
Both quantum mechanical cosmologies are governed by
the one and the same time-independent single-variable
Hartle-Hawking wave function.
It is only that its original argument, the cosmic scale
factor $a$, is replaced by $a\phi$ ($\phi$ being the
dilaton field) to form a Dirac in-scalar.
The Weyl vector enters quantum cosmology only
in the presence of an extra dimension, where its fifth
component, serving as a 4-dim Kaluza-Klein in-scalar,
governs the near Big Bang behavior of the wave function.
The case of a constant Kaluza-Klein in-radius is
discussed in some detail.

\end{abstract}

%\pacs{}

\maketitle

\noindent \textbf{\textit{Introduction}}\smallskip

The mini-superspace approximation \cite{mini}, while
being premeditatedly naive and simple by construction,
is still one of the best available theoretical tools to probe
the quantum cosmology. 
The prototype mini-superspace Hartle-Hawking model
is economical in its ingredients. They include:
(i) The Einstein-Hilbert action,
(ii) A positive cosmological constant $\Lambda>0$, and
(iii) A spatially open Universe $\kappa>0$.
The resulting wave function $\psi(a)$, strikingly
time-independent reflecting the Hamiltonian constraint,
obeys the Wheeler-deWitt equation
(reduced to a zero energy Schrodinger equation),
but can still have different spontaneous creation
interpretations
(Hartle-Hawking \cite{HH}, Linde \cite{Linde},  Vilenkin
\cite {Vilenkin}) differing from each other by the initial
conditions.
Renewed interest in the so-called Hartle-Hawking
no-boundary proposal has been reported \cite{recentmini},
with the presumably problematic Euclidean approach
\cite{Euclid} being replaced by the potentially tenable
Lorentzian approach \cite{Lorentz} alternative.

But is the exact $\psi_{HH}(a)$ a unique fingerprint of
the underlying theory of general relativity?
We first prove that the answer to this question is
negative.
One cannot tell, at least at the mini-superspace level,
the Einstein-Hilbert gravity from the Weyl-Dirac gravity
\cite{WD} (not to be confused with $C^2$-gravity
\cite{C2}, where $C_{\mu\nu\lambda\sigma}$ stands
for the Weyl tensor).
Furthermore, the conclusion holds for an arbitrary
Brans-Dicke parameter \cite{BD}.

This brings us to the more general topic of no-scale
quantum cosmology, where the underlying gravitational
theory exhibits local scale invariance.
The latter local symmetry is translated into an additional
constraint (on top of the Hamiltonian constraint), and may
have a far reaching impact on the mini-superspace.
The associated no-scale wave function of the Universe
can only depend, using Dirac language, on in-scalars.
Unfortunately, the no-scale $C^2$ conformal cosmology
is empty, leaving the stage, as already mentioned earlier, to
Weyl-Dirac cosmology and two scalar gravity-anti-gravity
cosmology \cite{GaG}.
For more complicated cases, one may invoke Kaluza-Klein
reduced higher dimensional local scale symmetric theories
\cite{conKK}.
This allows the Weyl vector to enter the game, and
even govern the wave function behavior \cite{conBB}
near Big Bang.

\bigskip
\noindent \textbf{\textit{Weyl-Dirac preliminaries}}
\smallskip

Let our starting point be the Brans-Dicke theory,
supplemented by a quartic potential for the dilaton
scalar field $\phi(x)$, described by the action
\begin{equation}
	{\cal I}_{BD}=\int d^4 x\sqrt{-g}\left(\phi^2 {\cal R}
	-4\omega g^{\mu\nu}\phi_{;\mu}\phi_{;\nu}
	-2\Lambda\phi^4\right) ~.
	\label{BD}
\end{equation}
The theory enjoys \emph{\textbf{global}} scale invariance
($\Omega=$ arbitrary constant) under
\begin{equation}
	g_{\mu\nu} \rightarrow
	e^{2\Omega}g_{\mu\nu}~,~~
	\phi \rightarrow  e^{-\Omega}\phi ~,
	\label{Omega}
\end{equation}
and furthermore exhibits the much reacher
\emph{\textbf{local}} scale invariance ($\Omega(x)=$
arbitrary function) for the critical case of
$\omega=-\frac{3}{2}$.
However, as prescribed by Dirac, local scale invariance
can be extended to accompany any Brans-Dicke
$\omega$-theory.
The corresponding Weyl-Dirac gravity is field theoretically
formulated by the action 
\begin{equation}
\begin{array}{ccc}
	&{\cal I}=
	\int d^4 x \sqrt{-g} \left(\phi^2 {\cal R}^\star
	-4\omega g^{\mu\nu}\phi_{\star\mu}\phi_{\star\nu}
	\right.   &  \vspace{4pt} \\
	&\displaystyle{-\frac{1}{4}g^{\mu\nu}g^{\lambda\sigma}
	K_{\mu\lambda}K_{\nu\sigma}
	-2\Lambda \phi^4)} ~.&
	\end{array}
	\label{WD}
\end{equation}
The Ricci scalar ${\cal R}$, known to govern the
Einstein-Hilbert action, has been consistently
supplemented by two terms which involve the Weyl
vector field $K^{\mu}$ and its divergence, such that
the generalized (stared) curvature
\begin{equation}
	{\cal R}^\star= 
	{\cal R}+g^{\mu\nu}\left(6K_{\mu;\nu}
	-6K_{\mu}K_{\nu}\right)
\end{equation}
transforms as a co-scalar under the local scale
symmetry, that is
${\cal R}^\star \rightarrow e^{-2\Omega(x)}{\cal R}^\star$.
Respectively, the (stared) co-covariant dilaton derivative,
defined by
\begin{equation}
	\phi_{\star \mu}=\phi_{; \mu}+K_{\mu}\phi ~,
\end{equation}
transforms ala
$\phi_{\star \mu}\rightarrow e^{-\Omega(x)}\phi_{\star \mu}$
in accord with Eq.(\ref{Omega}).
The last mandatory ingredient in the prescription is
of course the Weyl Maxwell-like gauge transformation
\begin{equation}
	K_{\mu}\rightarrow K_{\mu}+\Omega_{;\mu}~,
\end{equation}
which leaves invariant the associated anti-symmetric
co-tensor $K_{\mu\nu}=K_{\mu \star\nu}
-K_{\nu\star\mu} =K_{\mu;\nu}-K_{\nu;\mu}$.

\bigskip
\noindent \textbf{\textit{No-scale quantum cosmology}}
\smallskip

To keep track of the Hartle-Hawking model, we
stick to an open universe ($\kappa>0$), and a positive
cosmological constant ($\Lambda>0$) in the Einstein
frame.
The cosmological FRWL line element
\begin{equation}
	ds^2=-n^2(t)dt^2+a^2(t)
	\left(\frac{dr^2}{1-\kappa r^2}
	+r^2 d\Omega^2\right)
\end{equation}
is accompanied by $K_{\mu}dx^\mu=v(t)dt$
and $\phi(t)$.

At the mini-superspace level, and for an arbitrary
Brans-Dicke parameter, the Weyl-Dirac action
Eq.(\ref{WD}) reduces after spatial integration to
$\int {\cal L}dt$, with the mini-Lagrangian taking
the form
\begin{equation}
	\begin{array}{c}
	{\cal{L}}= na(6\kappa-2\Lambda a^2\phi^2)\phi^2
	\vspace{5pt}\\
	\displaystyle{+\frac{2a}{n}\left(
	(3+2\omega)a^2(v\phi+\phi^\prime)
	-3(\phi^{\prime} a+\phi a^{\prime})^2 \right)~,}
	\end{array}
	\label{miniL}
\end{equation}
to be compared with the corresponding Hartle-Hawking
mini-Lagrangian
\begin{equation}
	{\cal{L}} _{HH}= n a (6\kappa-2\Lambda a^2) -
	\frac{6a}{n}a^{\prime 2}~.
\end{equation}
The canonical variables $n$ and $v$ are non-dynamical, and as
such one is not allowed to prefix (= fix at the level
of the Lagrangian) their values.
Prefixing $n$ would kill the Hamiltonian constraint,
and by the same token, prefixing $v$ means throwing
away the scale invariance constraint.
To be a bit more specific, the two associated primary
constraints are
\begin{equation}
	p_v=\frac{\partial \cal{L}}{\partial \phi^\prime}
	\approx 0 ~,~~
	p_n=\frac{\partial \cal{L}}{\partial n^\prime}
	\approx 0~, 
	\label{constraints}
\end{equation}
they are clearly first class, exhibiting vanishing Poisson
brackets
\begin{equation}
	\{p_n , p_v\}=0 ~.
\end{equation}
The two left over non-trivial momenta are given by
\begin{eqnarray}
	&& p_a=-\frac{12a\phi}{n}(a\phi)^\prime ~,\\
	&& p_\phi=\frac{4a^2}{n}
	\left((3+2\omega)a(v\phi+\phi^\prime)
	-3(a\phi)^\prime\right) ~.
\end{eqnarray}
For non-critical scale invariance, that is 
$3+2\omega\neq 0$, one can now inversely calculate
the velocities
\begin{eqnarray}
	&&a^\prime= va- n\frac{3\phi p_{\phi}
	+2\omega a p_a}{12(3+2\omega)a^2\phi^2}~, \\
	&& \phi^\prime=-v\phi+n\frac{\phi p_\phi-a p_a}
	{4(3+2\omega)a^3\phi}~,
\end{eqnarray}
to be substituted into
${\cal H}=\sum p\dot{q}-{\cal L}$,
and eventually arrive at the Weyl-Dirac mini-superspace
Hamiltonian
\begin{equation}
	\begin{array}{c}
	 {\cal H}=v(a p_a-\phi p_\phi) \vspace{5pt}\\  
	\displaystyle{-\frac{n}{a}\left(  
	\frac{p_a^2}{24\phi^2}-
	\frac{(a p_a-\phi p_\phi)^2}{8(3+2\omega)a^2\phi^2}
	+2(3\kappa-\Lambda a^2 \phi^2)a^2\phi^2\right)}~.
	\end{array}
	\label{HWD}
\end{equation}
It is notably linear in $n,v$, and should be compared
with its Hartle-Hawking analogue
\begin{equation}
	{\cal H}_{HH}=-\frac{n}{a}
	\left(\frac{p_a^2}{24}+
	6\kappa a^2-2\Lambda a^4\right) ~.
\end{equation}
On self consistency grounds, the two associated first class
constraints Eq.(\ref{constraints}) must Poisson
commute with the mini-Hamiltonian Eq.(\ref{HWD}),
thereby leading to
\begin{eqnarray}
	&& \{p_v,{\cal H}\}=\phi p_\phi-a p_a=0 ~, \\
	&& \{p_n,{\cal H}\}=\frac{1}{a}\left(
	\frac{p_a^2}{24\phi^2}+
	6\kappa a^2 \phi^2-2\Lambda a^4 \phi^4\right)=0 ~.
\end{eqnarray}
These equations must be satisfied not only classically,
but following Dirac prescription, quantum mechanically
as well, thus giving rise to two Schrodinger
Wheeler-deWitt equations for the time-independent
quantum mechanical wave function $\psi(a,\phi)$.

Let $p=-i \hbar \frac{\partial}{\partial q}$, and as
usual replace $qp$ by its symmetrized Hermitian 
version $\frac{1}{2}(qp+pq)=qp-\frac{1}{2}i\hbar$.
Consequently, the first Schrodinger equation, reflecting
the scale invariance constraint, takes the
$\hbar$-independent form
\begin{equation}
	\left(a\frac{\partial}{\partial a}
	-\phi\frac{\partial}{\partial \phi}\right)
	\psi(a,\phi)=0~,
\end{equation}
whose most general solution is given by
\begin{equation}
	\psi(a,\phi)=\psi(a\phi)~,
\end{equation}
meaning a single variable wave function.
This is of course not a coincidence.
Recalling the local scale symmetries of the
mini-Lagrangian, in particular
$a(t)\rightarrow e^{\Omega}a(t)$ along with
$\phi(t)\rightarrow e^{-\Omega}\phi(t)$, so that
the product $b(t)=a(t)\phi(t)$ transforms like
an in-scalar.
This is just a simple example of the general rule: 
\emph{In no-scale mini-superspace cosmology, the wave
function $\psi$ can only depend on in-tensors,
carrying no length units}.

The second Schrodinger equation, reflecting the
Hamiltonian constraint, then becomes
\begin{equation}
	-\frac{\hbar^2}{24}
	\frac{d^2 \psi(b)}{d b^2}
	+(6\kappa b^2
	-2\Lambda b^4)\psi(b)=0~.
	\label{HH}
\end{equation}
This is now an ordinary differential equation, recognized
as the original Hartle-Hawking equation for the
in-scalar
\begin{equation}
	b=a\phi ~.
\end{equation}
This simply means that, at the mini superspace level, one
cannot really tell traditional Einstein-Hilbert gravity
from the local scale invariant Weyl-Dirac gravity. 
In other words, the wave function $\psi_{HH}(a)$
is traded for $\psi_{HH}(a\phi)$.
The result being $\omega$-independent.

\bigskip
\noindent \textbf{\textit{Weyl-Dirac Kaluza-Klein reduction}}
\smallskip

To introduce more ingredients into the no-scale cosmology,
we consider a Kaluza-Klein reduction of Weyl-Dirac gravity
in 5-dimensions, given explicitly by
\begin{equation}
\begin{array}{ccc}
	&\displaystyle{{\cal I}_5=
	\int d^5 x \sqrt{-G} \left(\phi^2 {\cal R}^\star_5
	-4\omega_5 G^{MN}\phi_{\star M}\phi_{\star N}
	\right. }  &  \vspace{4pt} \\
	&\displaystyle{~~-\frac{1}
	{4}\phi^{\frac{2}{3}}G^{MN}G^{PQ}
	W_{MP}W_{NQ}
	-2\Lambda \phi^{\frac{10}{3}}) ~.}&
	\end{array}
	\label{WD5}
\end{equation}
Here, adjusting the coefficients and the various powers
of $\phi$ to fit the 5-dim world, we also have
\begin{eqnarray}
	& {\cal R}^\star_5={\cal R}_5
	 +G^{MN}\left( 8W_{M;N} 
	-12W_{M}W_{N}\right) ~, & \\
	& \displaystyle{\phi_{\star M}=
	 \phi_{; M}+\frac{3}{2}W_M \phi} ~,&
\end{eqnarray}
where $W_M$ stands here for the 5-dim Weyl vector, and
correspondingly $W_{MN}=W_{M;N}-W_{N;M}$.
We further note that, in 5-dimensions, the critical value
of the Brans-Dicke parameter is $\omega_5=-\frac{4}{3}$.

Truly, Kaluza-Klein compactification does introduce
a fundamental length scale into the theory, viz.
\begin{equation}
	dx_5=\ell d\theta  \quad(\Delta\theta=2\pi) ~.
\end{equation}
However, owing to the fact that $x_5$-independence is
imposed on a classical scale invariant theory, the Kaluza-Klein
radius $\ell$ can be fully absorbed within the redefinitions
of the fields which constitute the 4-dim effective theory.
It is to say that after integrating out the 5-th dimension, the
effective 4-dim ground state stays locally scale symmetric
(in the 4-dim language).

Carrying out the Kaluza-Klein reduction procedure, the
5-dim line element dictionary reads
\begin{equation}
	ds_5^2=S^{-\frac{1}{3}}ds_4^2
	+S^{\frac{2}{3}}\ell^2(d\theta+A_\mu dx^\mu)^2 ~,
	\label{ds5}
\end{equation}
with the algebraic advantage that
$R_5 \sqrt{-G}=R_4 \sqrt{-g}+...$
Similarly, the 5-dim Weyl vector field can be
decomposed into
\begin{equation}
	W_M dx^M=
	\frac{2}{3}K_\mu dx^\mu
	+s\ell(d\theta+A_\mu dx^\mu)~,
\end{equation}
where the two 4-dim vector fields, the Weyl vector
field $K_\mu$ and the Maxwell vector field $A_\mu$,
have been normalized and fully diagonalized (imitating
a $U(1)\otimes \tilde U(1)$ gauge theory).
To this we add the simple yet powerful relation
\begin{equation}
	\omega_5=\omega_4+\frac{1}{6} ~,
\end{equation}
so that $\frac{4}{3}(4+3\omega_5)=2(3+2\omega_4)$,
thereby assuring that the critical 5-dim Brans-Dicke
theory properly reduces, as it should, down to the critical
4-dim Brans-Dicke theory.

After integrating out the Kaluza-Klein circle, that is
$\int {\cal L}_5d^5 x \rightarrow
2\pi\ell \int {\cal L}_4 d^4 x$, it becomes
mandatory to redefine the various scalars floating
around by adjusting their length units to fit the 4-dim
language.
Indeed, uniquely redefining according to
\begin{equation}
\begin{array}{l}
	\phi \mapsto \ell^{-\frac{1}{2}} \phi ~ , \vspace{4pt} \\
	S \mapsto  \ell^{-2} S ~ , \vspace{4pt} \\
	s \mapsto  \ell^{-1} s ~ , 
\end{array}
\end{equation}
will leave the action $\ell$-independent, meaning that
local scale invariance has been restored.
The interplay among these scalars, reflecting their
specific charges under scale symmetry, will be discussed
soon.
Crucial for the forthcoming discussion is the identification
of the in-scalars involved.

Altogether, the effective 4-dim theory is described by
the Lagrangian
\begin{equation}
	\begin{array}{ccc}
	\displaystyle{\frac{{\cal L}}{\sqrt{-g}}=\phi^2 R^\star_4
	-4\omega_4g^{\mu\nu}\phi_{\star\mu}\phi_{\star\nu}}
	-\frac{1}{4}\phi^2 S F^{\mu\nu}F_{\mu\nu}
	 \vspace{4pt}\\ 
	 \displaystyle{-\frac{1}{4}(\phi^2 S)^{\frac{1}{3}}
	 \left(\frac{2}{3}K^{\mu\nu}+sF^{\mu\nu}\right) 
	 \left(\frac{2}{3}K_{\mu\nu}+sF_{\mu\nu}\right)}
	 \vspace{4pt}\\ 
	 \displaystyle{
	 -\frac{1}{6}\phi^2 g^{\mu\nu}
	 \frac{(\phi^2 S)_{\star\mu}}{\phi^2 S}
	 \frac{(\phi^2 S)_{\star\nu}}{\phi^2 S}
	 -\frac{\phi^2}{(\phi^2 S)^{\frac{2}{3}}}
	 g^{\mu\nu}s_{\star\mu}s_{\star\nu}}
	 \vspace{4pt}\\ 
	 \displaystyle{-\left(\frac{9}{2}
	 (3+2\omega_4)\frac{s^2}{\phi^2 S}
	+\frac{2\Lambda}
	{(\phi^2 S)^{\frac{1}{3}}}\right)\phi^4} ~,
	\end{array}
	\label{effL}
\end{equation}
where the 4-dim star derivatives are given explicitly by
\begin{equation}
\begin{array}{l}
	\phi_{\star\mu}
	= \phi_{;\mu}+K_{\mu} \phi ~ , \vspace{4pt} \\
	(\phi^2 S)_{\star\mu}
	= (\phi^2 S)_{;\mu} ~ , \vspace{4pt} \\
	s _{\star\mu} = s_{;\mu} ~. 
\end{array}
\end{equation}
The detailed derivation has been carried out elsewhere,
but in any case, Eq.(\ref{effL}) can have life of its own.
While the above 4-dim Lagrangian may look a bit messy,
it is nevertheless invariant under the local Maxwell
gauge transformations
\begin{equation}
	A_\mu\rightarrow A_\mu+\chi_{;\mu} ~,~
	K_\mu\rightarrow K_\mu ~,
\end{equation}
and in particular under the 4-dim local Weyl scale
transformations
\begin{equation}
\begin{array}{ccc}
	& K_\mu\rightarrow K_\mu+\Omega_{;\mu} ~,~
	A_\mu\rightarrow A_\mu ~, & \vspace{6pt}\\
	& g_{\mu\nu} \rightarrow e^{2\Omega} g_{\mu\nu},~
	\phi \rightarrow e^{-\Omega}\phi ,~
	S \rightarrow e^{2\Omega} S ,~
	s \rightarrow s~.&
	\end{array}
\end{equation}
We can thus finally identify the in-scalars of the theory,
relevant for quantum cosmology.
They are the built-in $s$, and the product $\phi^2 S$.
In turn, any 'ordinary' 4-dim scalar field, having conventional
units of $(length)^{-1}$, must be of the form
$\phi s^p (\phi^2 S)^q$ for arbitrary $p,q$ .

Special attention should be devoted to the so-called
Maxwell-Weyl kinetic mixing \cite{mix} $s K^{\mu\nu}F_{\mu\nu}$
mediated by the in-scalar $s$.
This is a unique feature which characterizes the
Weyl-Dirac Kaluza-Klein interplay.
Unfortunately, it does not play a direct role in no-scale
cosmology, and as such, its remarkable aspects will be
discussed in detail in a sequel publication.

\bigskip
\noindent \textbf{\textit{No-scale Kaluza-Klein quantum
cosmology}}
\smallskip

At the 4-dim mini-superspace level, cosmology can only
tolerate the pure gauge configurations
\begin{equation}
	A_{\mu}=(A(t),0,0,0) ~,~~  K_{\mu}=(v(t),0,0,0) ~,
\end{equation}
for which $F_{\mu\nu}=K_{\mu\nu}=0$.
While the various fields are neutral under the Kaluza-Klein
$U(1)$ symmetry, and as such do not couple to $A_{\mu}$,
the Weyl vector field $K_{\mu}$ does enter the game via
the star derivatives of the scalar fields.

It takes then some algebra, a bit lengthy but straightforward, 
to translate the mini-superspace version of the Lagrangian
Eq.(\ref{effL}) into the Hamiltonian formalism.
The result
\begin{equation}
	\begin{array}{c}
	 {\cal H}=v(a p_a-\phi p_\phi+2S p_S) \vspace{5pt}\\  
	\displaystyle{-\frac{n}{a}\left[ 
	\frac{p_a^2}{24\phi^2}
	-\frac{3S^2 p_S^2}{2a^2\phi^2}
	-\frac{S^{\frac{2}{3}}p_s^2}{2a^2 \phi^{\frac{2}{3}}}
	-\frac{(a p_a-\phi p_\phi+2S p_S)^2}{8(3+2\omega)a^2\phi^2}
	\right.}  \vspace{5pt}\\  
	\displaystyle{\left. +6\kappa a^2\phi^2
	-\left(\frac{9(3+2\omega_4)s^2}{2S\phi^2}
	+\frac{2\Lambda}{S^{\frac{1}{3}}\phi^{\frac{2}{3}}}
	\right)a^4\phi^4\right]}
	\end{array}
	\label{HKK}
\end{equation}
is to be compared, term by term, with the simpler 4-dim
Hamiltonian Eq.(\ref{HWD}).
As before, we face a Hamiltonian linear in $v ,n$, giving
rise to two first class constraint.
Since our interest primarily lies with quantum no-scale
cosmology, we momentarily skip the classical equations
of motion and their solutions, and proceed directly to
the pair of Schrodinger equations.
The coefficient of $v$ in Eq.(\ref{HKK}) is immediately
recognized as  the $\hbar$-independent scale symmetry
constraint, leading to
\begin{equation}
	\left(a\frac{\partial}{\partial a}
	-\phi\frac{\partial}{\partial \phi}
	+2 S\frac{\partial}{\partial S}
	\right)\psi(a,\phi,S,s)=0 ~.
\end{equation}
This leaves the wave function to solely depend on
Dirac's in-scalars, e.g. on 
\begin{equation}
	a^\alpha \phi^\beta S^\gamma
	~~\quad \text{for}~ \alpha-\beta+2\gamma=0 ~.
\end{equation}
Without losing generality, the simplest choice
would be
\begin{equation}
	\psi(a,\phi,S,s)=\psi(b,z,s) ~,
\end{equation}
where we have used the short hand in-scalar notations
\begin{equation}
	b=a\phi~,~~ z=\log S \phi^2~.
\end{equation}

The Associated Hamiltonian constraint, identified as the
coefficient of $n$ in Eq.(\ref{HKK}), eventually becomes
the zero energy Schrodinger equation
\begin{equation}
	-\frac{\hbar^2}{24}
	\frac{\partial^2 \psi}{\partial b^2}
	+\frac{3\hbar^2}{2b^2}
	\frac{\partial^2 \psi}{\partial z^2}
	+\frac{e^{\frac{2}{3}z}\hbar^2}{2b^2}
	\frac{\partial^2 \psi}{\partial s^2}
	+V(b,z,s)\psi=0 ~,
\end{equation}
with the accompanying potential being
\begin{equation}
	V(b,z,s)=6\kappa b^2
	-\left(\frac{9}{2}(3+2\omega_4)e^{-z}s^2
	+2e^{-\frac{1}{3}z}\Lambda \right)b^4 ~.
\end{equation}
The door is now widely open for a variety of spacial
cases.
Of particular interest is perhaps the $\Lambda=0$
case, where the coefficient of $b^4$ can still be positive
provided $3+2\omega_4>0$.
In some respects, the $(3+2\omega_4)s^2$-term resembles
then the role of $\Lambda_{eff}(s)$, and may even be tiny if
the wave function is somehow concentrated around $s^2\ll1$
(this point will be sharpened soon).
At any rate, on simplicity grounds, we choose to analyze in
some detail only the special case of a constant Kaluza-Klein
in-radius.

%% Fig psi %%%
\begin{figure}[b]
	\center
	\includegraphics[scale=0.5]{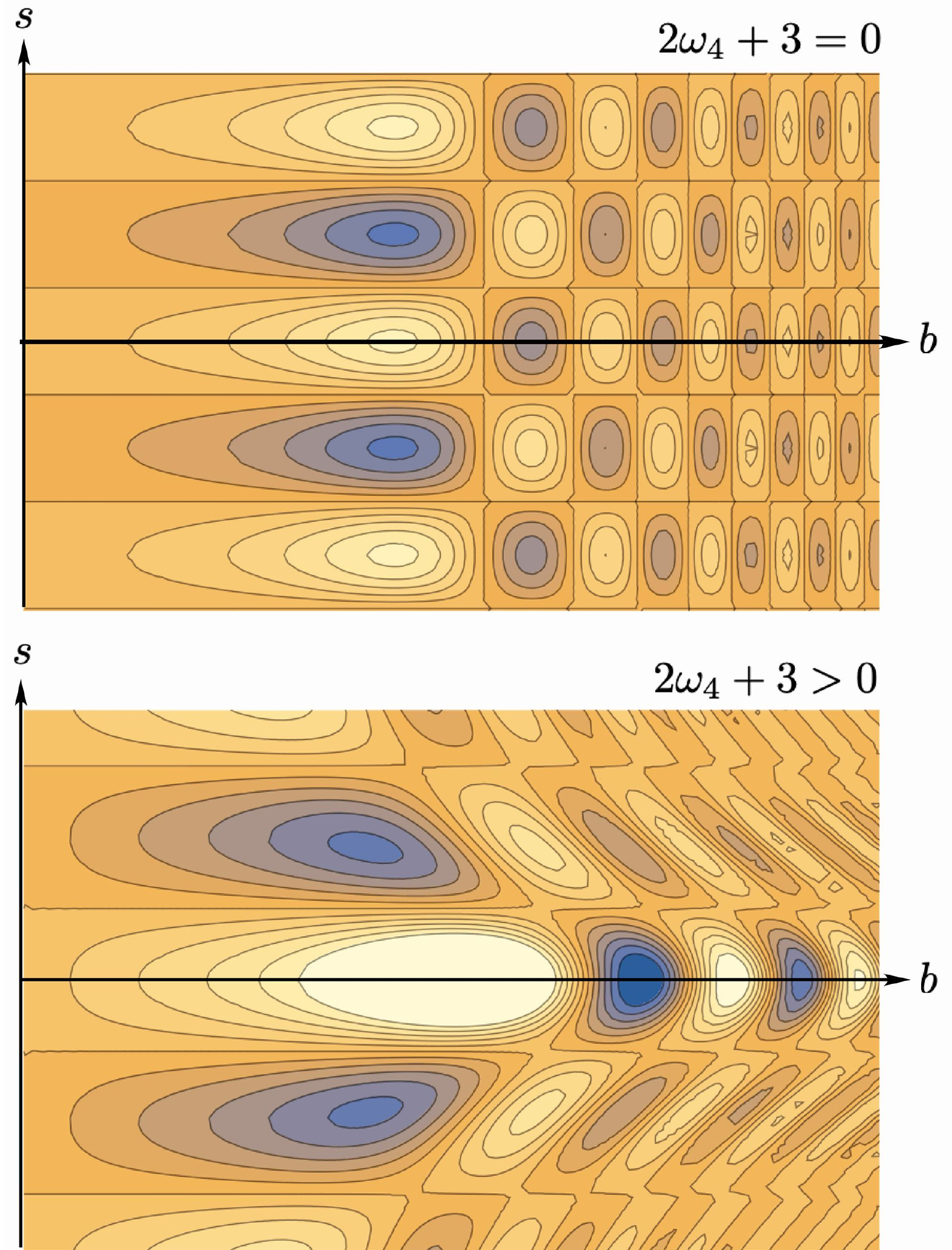}
	\caption{Contour Plots of the no-scale cosmological
	wave function $\psi(b,s)$, subject to the enforced/automatic
	deWitt initial condition $\psi(0,s)=0$.
	The critical Brans-Dicke special case
	$\omega_4=-\frac{3}{2}$ (upper plot), plotted for $\eta<0$,
	highly reminds us (up to $s$-periodicity) of the
	Hartle-Hawking solution.
	The super-critical case $\omega_4>-\frac{3}{2}$ (lower
	plot), capable of surviving the $\Lambda\rightarrow 0$
	limit, favors small values of the in-scalar $s$.}
	\label{psi}
\end{figure}
%%%%%%

\bigskip
\noindent \textbf{\textit{Constant Kaluza-Klein in-radius}}
\smallskip

In the standard Kaluza-Klein theory, with the line
element Eq.(\ref{ds5}), the invariant 5-dim radius is
given by $S^{\frac{1}{3}}\ell$.
In the earliest versions of the theory (the original
works of Kaluza and Klein, separately), the scalar degree
of freedom was in fact frozen $S=1$.
Such an ansatz, while quite welcome on mathematical
simplicity grounds, does not make any sense once local
scale symmetry is applied.
The closest one can get is by freezing a tenable in-scalar
(nothing to do with gauge fixing), with the obvious in-ansatz
being
\begin{equation}
	S \phi^2=1 ~.
\end{equation}
This will allow us to focus on the special role (beyond
Weyl-Maxwell mixing, which anyhow does not have
any cosmological fingerprints) played by the Weyl 4-dim
in-scalar $s$.
The corresponding 'handicapped' wave function
$\psi(b,s)$ obeys the Hartle-Hawking equation
\begin{eqnarray}
	&\displaystyle{-\frac{\hbar^2}{24}
	\frac{\partial^2 \psi}{\partial b^2}
	+\frac{\hbar^2}{2b^2}
	\frac{\partial^2 \psi}{\partial s^2}
	+V(b,s)\psi=0 ~,}& \\
	& \displaystyle{V(b,s)=6\kappa b^2
	-\left(\frac{9}{2}(3+2\omega_4)s^2
	+2\Lambda \right)b^4} ~.&
\end{eqnarray}

The critical case $\omega_4=-\frac{3}{2}$ is the easiest
to handle.
It allows for the separation of variables $\psi(b,s)=f(b)g(s)$,
with $f(b)$ serving as a modified Hartle-Hawking wave
function subject to the effective potential
\begin{equation}
	V_{eff}(b)=\frac{\eta\hbar^2}{2b^2}
	+6\kappa b^2-2\Lambda b^4~.
\end{equation}
and where the constant $\eta$ governs the equation
\begin{equation}
	g^{\prime\prime}(s)=\eta g(s) ~.
\end{equation}
Depending on the sign of $\eta$, new quantum
phenomena is expected to arise near the Big Bang
origin $b\rightarrow 0$.
The three cases are: 

\noindent (i) If $\eta=0$, the Hartle-Hawking model
is fully recovered, accompanied by $g(s)=const$.
The no-boundary proposal recovered for $f(b)\sim b$.

\noindent (ii) If $\eta>0$ then $g(s)=e^{\pm\sqrt{\eta}s}$
is unbounded, presumably signaling a non-physical case.

\noindent (iii) If $\eta<0$ then
$g(s)=e^{\pm i \sqrt{|\eta|}s}$ is well behaved.
If furthermore $\eta\hbar^2 \Lambda^2+16\kappa^3>0$,
as evident from the shape acquired by the effective
potential in this case, cosmic evolution undergoes a
(classically allowed) embryonic era.
The no-boundary proposal is not recovered.
However, the bonus in this case is automatic deWitt
initial conditions, as both solutions
$f_{1,2}(b)\sim b^{\delta_{1,2}}$ (with
$0<Re(\delta_{1,2})<1$) vanish asymptotically at the
origin.
For further details, see Fig.(\ref{psi}), upper plot.

The non-critical case, for comparison, is characterized
by an effective cosmological constant
\begin{equation}
	\Lambda_{eff}(s)
	=\Lambda+\frac{9}{4}(3+2\omega_4)s^2 ~.
\end{equation}
The supplemented term is positive for a super-critical 
Brans-Dicke parameter $\omega_4 >-\frac{3}{2}$
(including, in particular, the ghost-free case
$\omega_4 \geq 0$).
While the separation of variables method does not work
any more, the structure of the Schrodinger equation is such
that the behavior of the wave function near the Big Bang
is not sensitive to the value of $\omega_4$.
The larger $\omega_4$, the more concentrated is the
wave function around $s^2\ll 1$.
For further details, see Fig.(\ref{psi}), lower plot.

\smallskip
\noindent \textbf{\textit{Acknowledgments}}
%\acknowledgments

{We cordially thank Michael Lublinsky for his kind support.
Tomer Ygael was supported by the Israel Science Foundation
grant $\# 1635/16$.}

\end{document}